\documentstyle[prl,aps]{revtex}

\begin{document}
\draft

\title{A magnetic field diagnostic for sonoluminescence}

\author{Tom Chou$^{1}$ and Eric G. Blackman$^{2}$}

\address{$^{1}$LASSP, Cornell University, Ithaca, NY 14853 \\
$^{2}$Institute of Astronomy, Cambridge University, 
Madingley Road, Cambridge CB3 OHA, England}

\date{\today}
\maketitle

\begin{abstract} 
This study is motivated by the extraordinary process of
single bubble sonoluminescence (SBSL), where an acoustically driven
spherical shock is thought to power the emitted radiation.
We propose new experiments using an external magnetic field
which can induce anisotropies in both 
the shock propagation and radiation
pattern. The effects will depend on the temperature,
conductivity, and size of the radiating region.  
Our predictions suggest that such a laboratory 
experiment could serve as an important diagnostic 
in placing bounds on these parameters and
understanding the physics of sonoluminescence.
\end{abstract} 

\pacs{}


\widetext

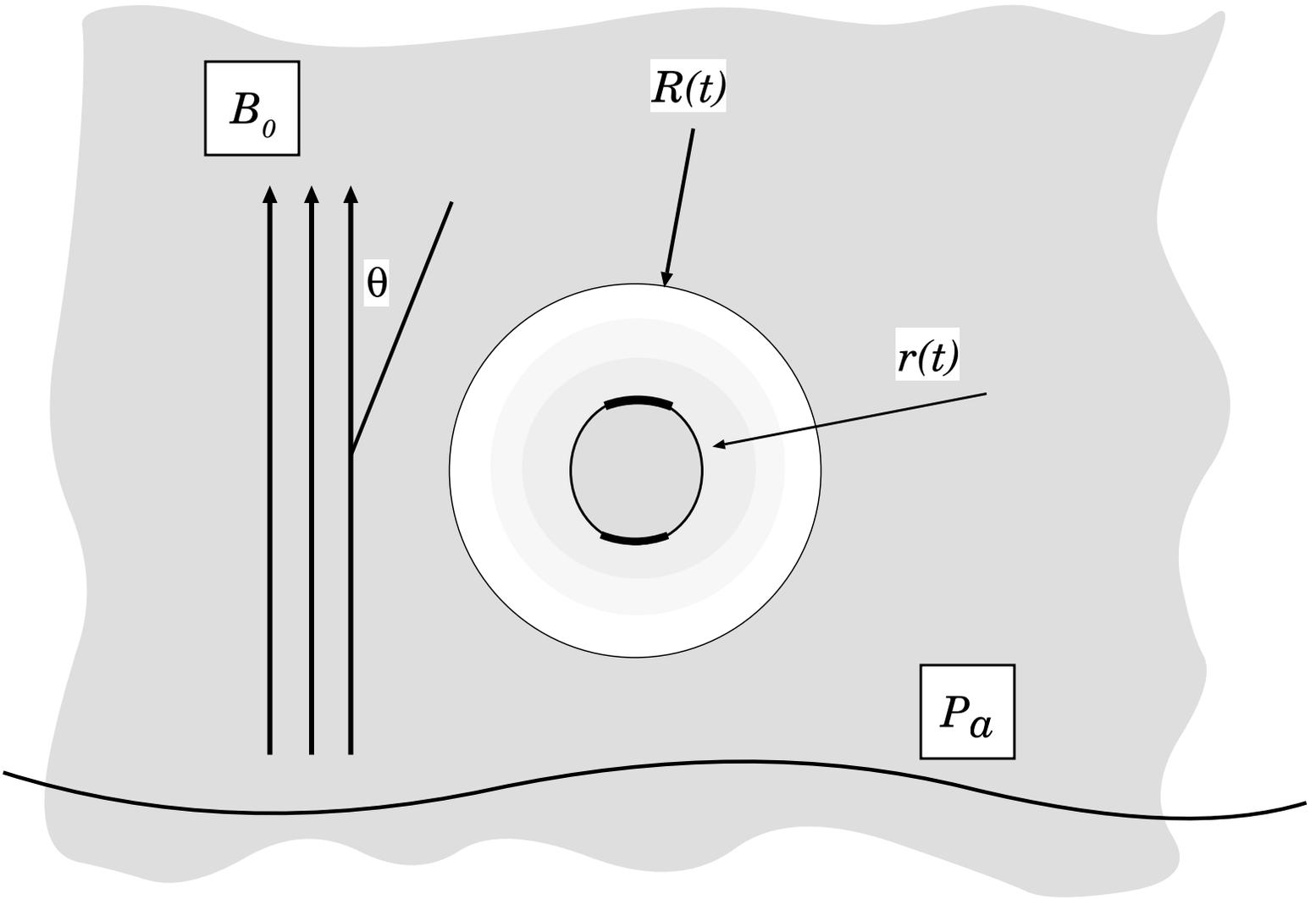
\begin{figure}
\caption{The sonoluminescing bubble in an external magnetic field
$B_{0}\hat{z}$. 
The liquid/gas interface and the shock front are denoted by $R(t)$ and 
$r(t)$ respectively. The external sound field supplies the pressure $P_{a}$.}
\label{BUBBLE}
\end{figure}

Sonoluminescence, first discovered in 1933 \cite{FIRST,WR}, 
is a remarkable phenomenon whereby sound is converted into
light. Recently, the process involving a single small gas bubble
trapped in a degassed liquid such as
water, which is then acoustically driven, has been 
studied \cite{PUTT}. Since the sound
wavelength is much larger than the bubble
radius $R(t)$, (see Fig. \ref{BUBBLE}), the bubble feels a
uniform pressure which varies with time. The
gas bubble then undergoes complicated nonlinear
oscillations.  At a threshold of the driving
pressure of about 1.15 atm, short ($ < 50$ ps),
intense pulses of light are radiated. The
acoustic energy of the system is thus focused into a
small region within the bubble for very short
times.  Recent interest in sonoluminesence has been in part motivated 
by possible technological applications in biophysics, sonochemistry and 
reactions at ultra-high temperatures \cite{SUSLICK}.

Experiments have probed the effects of ambient
temperature, liquid composition, gas composition, driving intensity, and
frequency on the phenomenon.  In
the experiments of Ref. \cite{PUTT}, there appears to
be a sharp transition from a nonradiating
oscillating bubble to a luminescing one.
Spectral fits to blackbody or Bremsstrahlung
emission are inconclusive but
suggest temperatures of at least 5000$^{\circ}$K 
\cite{PUTT,SUSLICK}.  

Theoretical attention has focused on
bubble dynamics, usually employing
variations of the Rayleigh-Plesset equation to model the 
bubble motion and stability \cite{PUTT,NADIM}.  
Furthermore, studies imply that as the gas bubble  
shrinks, it launches an inwardly 
propagating shock wave \cite{WU}.  
The shock collapses to the center, rebounds, and
can hit the liquid/gas bubble wall from which it was launched. 
Radiation is thought to be emitted immediately
after the shock rebounds from the center, where
extremely high temperatures and ionization are
predicted \cite{WU}. Numerical simulations 
suggest temperatures of $10^{8\,\circ}$K  
\cite{WU}, much higher than those implied by experiments. 
Thermal blackbody and/or Bremsstrahlung radiation,
Casimir effects, \cite{SCHWINGER} or decay of
excited molecular states, have all been suggested as
relevant processes involved in the luminescence, but
the emission spectra in SBSL are so featureless 
that distinguishing among radiation mechanisms is
difficult. Despite attention on the effects of
bubble dynamics, diffusion, and material composition
on sonoluminescence, our understanding of the 
underlying physical mechanisms   
remains an unsorted mixture of nonlinear
hydrodynamics, shock physics and chemical reaction kinetics.   
However, one emerging criterion for SBSL
is the necessity of stability of spherical symmetry 
in bubble structure $R(t)$\cite{BRENNER}.  The 
transient symmetric stability of the converging shock is
also though to be crucial for SBSL.
   
In this Letter, we suggest the use of an externally applied
magnetic field as a diagnostic by catalyzing the breaking
of spherical stability and putting  bounds on the 
temperature, conductivity, and ionization of 
the sonoluminescing bubble. 
The external magnetic field can
break spherical symmetry of the shock, bubble, or 
radiation patterns and disrupt or alter 
sonoluminescent behavior. 

{\it Effects of} $\vec{B}$ {\it on bubble structure - } 
We assume that the bubble gas is ionized at some point 
in the acoustic cycle. Adiabatic
compression of the gas bubble $R(t)$ is
probably sufficient to partially ionize the gas;
Saha's equation for temperatures
of 7000$^{\circ}$K give roughly 5$\%$ ionization.   
First consider the possibility that during the adiabatic collapse,
a spherical region of plasma has a high 
conductivity $\sigma_{p}$  such that 
the magnetic Reynolds number 
$R_{M} = \sigma_{p}vR^{*}\gg 1$ (where $v, R^{*}$ are 
typical velocities  and the radius of the ionized region). 
Alfv\'en's theorem of flux freezing holds, and the application 
of $\vec{B}_{ext}$ adds an magnetic stiffness in the 
direction perpendicular to $\vec{B}_{ext}$. The ionized region 
then  collapses asymmetrically, 
with the poles compressing more than the equator. 
If $R_M \gg 1$ is not achieved by initial adiabatic compression,
shocks launched by the liquid/gas interface
may further compress, heat, and ionize 
the plasma, consistent with  numerical simulations \cite{WU}.

Shocks form when the hydrodynamic velocity
exceeds a characteristic group velocity. 
In the presence of a magnetic field,
characteristic group velocities depend on the magnitude and 
orientation of the applied field. For $R_{M} \gg 1$, the MHD
modes have group velocities \cite{BOYD}

\begin{equation}
v^{2} = {1\over 2}(c_{s}^{2}+v_{A}^{2})\pm{1 \over
2}\sqrt{(c_{s}^{2}+v_{A}^{2})^{2}-4c_{s}^{2}v_{A}^{2}
\cos^{2}\theta}
\end{equation}

\noindent and $v = v_{A}\cos \theta$, where
$v_{A}^{2} = B^{2}/(4\pi\rho)$ and $c_{s}$ is the 
hydrodynamic sound speed ($B$ is the magnetic field 
in the region of interest and $\rho$ is the plasma density).  
These are the magneto-acoustic waves and
the Alfv\'en wave respectively.  
The angular dependence of these velocities
will break the spherical symmetry of a plasma shock.

We consider a shock just after rebound since here
the upstream region was previously
shocked during pre-rebound and more likely 
to be in a high $R_M$ state.  We will first
assume $R_M \gg 1$ both upstream and downstream. 
(The possibly important but rather involved 
case of ionizing shocks will not be treated here.)
The governing equations are the mass, momentum, 
and energy conservation, and the steady 
state Maxwell's equations and  Ohm's law,  
$\vec{E} \simeq c/(4\pi\sigma)\nabla\times\vec{B}
-(\vec{v}/c)\times\vec{B}$, valid for 
$\omega_{B}^{-1} = m_{e}c/eB_{0} <$
collision time ($c, m_{e}, e$ are the speed of light, 
electron mass, and electron charge respectively).  
Locally integrating the conservation laws
perpendicular to a shock, these 
equations are $\nabla\cdot\vec{B} =
\nabla\cdot(\rho\vec{v})=0$ and, \cite{MARSHALL}

\begin{equation}
\begin{array}{c}
\eta\,\hat{n}\times\nabla\times\vec{B} = 
\left[\hat{n}\times(\vec{v}\times\vec{B})\right]^{x}_{0}, \\
\rho\nu\left(\hat{n}\!\cdot\!\nabla v_{i} + {1 \over 3}
n_{i}\nabla\!\cdot\!\vec{v} \right) =
\left[(\vec{B}\!\cdot\!\vec{n})B_{i}/4\pi -\rho v_{i}(\vec{v}\!\cdot\!\vec{n})
-P_{M}n_{i}\right]^{x}_{0}, \\
{\eta \over 4}\hat{n}\!\cdot\!
(\nabla\!\times\!\vec{B})\!\times\!\vec{B} = \left[ \hat{n}\!\cdot\!\vec{v}(u
+{1 \over 2}\rho v^2 + P_{M})- {1 \over 4\pi}
(\vec{B}\!\cdot\!\vec{v})(\vec{B}\!\cdot\!\hat{n})\right]^{x}_{0},\\
\label{J1}
\end{array}
\end{equation}

\noindent where $[S]^{x}_{0} \equiv S(x)-S(0)$ and 
$u=P/(\gamma-1)\rho$ is the internal energy of the gas
($P, \eta$, and $\gamma = C_{p}/C_{v}$ are the pressure, shear viscosity
and adiabatic index of the gas, respectively).  
The limits straddling the shock are far outside the shock bubble 
(0), where all quantities are uniform, and $x$, a normal distance
downstream. The total pressure $P_{M}
\equiv  P+B^{2}/8\pi$ (for $R_{M} \gg 1$, the $E^{2}$ contribution is 
smaller than $B^2$ by $(v/c)^{2}$, which  is small; for $R_{M}\ll 1$,
$E^2$ is smaller by $(v/c)^{2}/R_M$). Equations (\ref{J1}) include dissipation due to 
plasma shear viscosity $\nu$ and resistivity $\eta \equiv c^2/(4\pi\sigma)$.
We can neglect the contribution of the any short intense radiation burst to
the electromagnetic stress tensor  if we
focus only on the shock conditions just before or after the 
radiative burst. Compression ratios and pressure jumps 
depend on $\theta$, the angle between the $\vec{B}_{0}$ and 
the shock normal. We thus expect a nonuniform MHD
shock forming first at the poles where $\theta = 0$ and 
$B_{0}$ offers no additional stiffness. 

In the high conductivity ($R_{M}\rightarrow \infty$) limit, 
all quantities vary over a thin shock front such 
that the fields at $x = 0^{-}$ are uniform 
and gradient terms vanish when we consider the 
quantities away from the sharp transition region.
Upon solving the resulting equations (Rankine-Hugoniot)  
we obtain,

\begin{equation}
\begin{array}{c}
\beta\left[\gamma M_{0}^{2}(1-1/s)-\Delta+1\right]
=(\alpha^{2}-1)\sin^{2}\theta \\
{2\gamma \over \gamma-1}\left({\Delta\over s}-1\right)
+\gamma M_{0}^{2}\left({1 \over s^{2}}-1\right)=
4\beta^{-1}(1-\alpha)\sin^{2}\theta -\gamma M_{0}^{2}\tan^{2}\theta 
\left(1-{\alpha \over s}\right)^{2}
\label{A1}
\end{array}
\end{equation}

\noindent where $\alpha \equiv  (\beta\gamma M_{0}^{2} - 2\cos^{2}\theta)/
(\beta \gamma M_{0}^{2}/s - 2\cos^{2}\theta), 
\beta = 8\pi P_{0}/B_{0}^{2}$ is the ratio of the  particle
to magnetic pressure, and $M_{0} = u_{0}/c_{s}$ is the Mach number. 
The  roots of Eqs. (\ref{A1}), $s$ and $\Delta\, (>1)$, 
determine the thermodynamically allowed shock states.

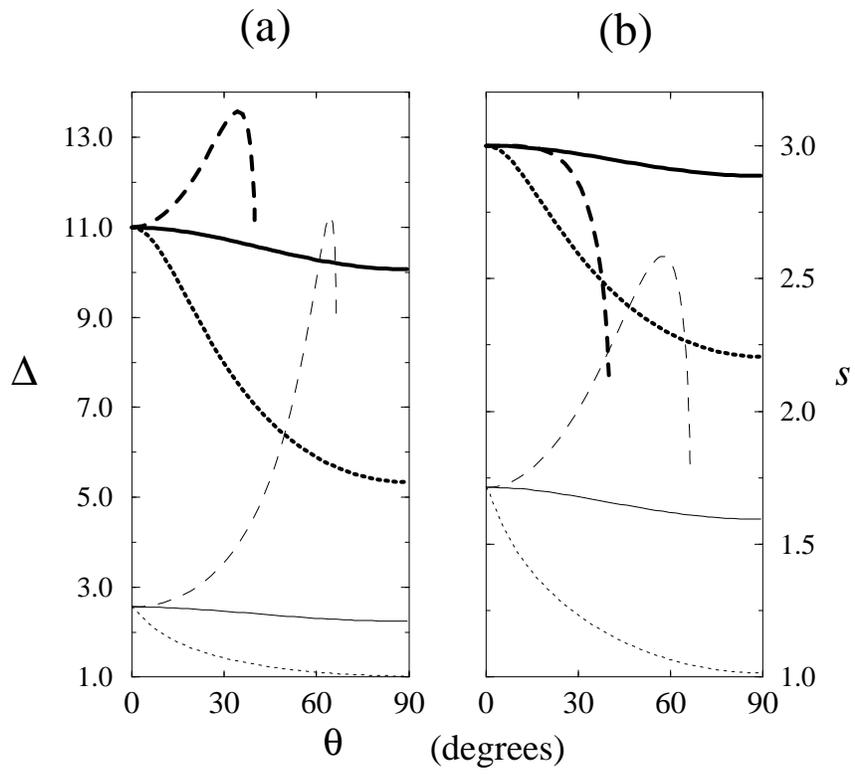
\begin{figure}
\caption{The pressure and density ratios for $R_{M} \gg 1$. (a)
$\Delta$ as a function of $\theta$ for $M_{0}=1.5$ (lower 3 thin curves),
and $M_{0}=3$ (upper 3 thick curves).  Solid, dotted, and dashed curves
correspond to $\beta = 10, 1, 0.1$, respectively. (b) Density ratio $s$ for similar 
values of $M_{0}$ and $\beta$.}
\label{ASYM}
\end{figure}

Fig. \ref{ASYM} shows the local shock structure for
$\gamma = 5/3$ and at varying $M_{0}$ and $\beta$. Except for 
low $\beta$ and $M_{0}$,  
the compression and density ratios,  $\Delta$ and
$s$, incur their greatest jumps at the poles. Numerically 
predicted pressures are on the order $10^{10}$
dyne/cm$^{2}$ \cite{WU}, thus, 
fields of $B_{0} \sim 10$T are required for $\beta \approx O(1)$. 
The angle $\theta$ can more or less be associated with the 
angle of $\vec{B}_{ext}$ in Fig. \ref{BUBBLE}. Deformations 
of the ionized region before shock formation will 
stretch the $x$-axes in Fig. \ref{ASYM} depending 
on $\beta$. Fig. \ref{ASYM}, however, shows the 
difference in jumps between $\theta=0, \pi/2$. 
For large $\beta$, the shock will be nearly spherical, 
but magnetic field compression near the equator would tend to 
decrease $\theta$ throughout the hemisphere. When $\beta$ is small,
the shock surface will be oblate owing to the lateral 
magnetic stiffness, and again, smaller $\theta$
will approximate most of the surface. In fact, the 
shock solution ends at a switch-off shock for low enough $\beta$. 
Dissipative processes will smear this region.

In the limit $R_{M}\rightarrow 0$ (low conductivity, or small velocity or length
scales) the magnetic field and the plasma flow 
are decoupled, and the plasma can slide freely through the field.
The discontinuities in $s, \Delta$ and $\vec{v}$  approach  those of a hydrodynamic shock
and the effect of the external $\vec{B}$-field for $R_{M} \ll 1$
is expected to be small. 
Though the  result given by (\ref{A1}) is correct 
for all $R_{M}$, for small $R_{M}$ the $\vec{B}$-field dissipation 
length $L$ becomes much  larger than the
viscous dissipation length  $\delta$ --
the scale over which pressure and velocity vary. 

Instead of solving the nonlinear differential 
equations (\ref{J1}) \cite{MARSHALL},
we  simply estimate the variation of $B$ across $\delta$
at $\theta=\pi/2$, where the largest  effect occurs.
Most of the variation in $\vec{v}, \rho$, and $P$ will occur
over $\delta$; far enough away from the shock the differences in the 
uniform states (0 and 1) are given by (\ref{A1}). 
Using $dB/dx \simeq (B_{0}-B_{1})/L$, and evaluating quantities 
in  Eqs. (\ref{J1}) across $\delta$, $s$ and $\Delta$ 
to order $R_{M}/\beta$ are



\begin{equation}
\begin{array}{c}
\displaystyle s=\bar{s} - 2{(M_{0}^2(\gamma-1)-\gamma)^2 
+M_{0}^2\gamma(\gamma-1) \over (M_{0}^2(\gamma-1) 
+\gamma)^{2}}
\left({R_{M}\over \beta }\right)f(\theta) \\
\displaystyle \Delta = \bar{\Delta} -
{2\gamma \over M_{0}^{2}}\left({4M_{0}^2\gamma \over \gamma-1} 
- {3\gamma+1 \over \gamma-1}\right)
\left({R_{M}\over \beta}\right)g(\theta),
\end{array}
\end{equation}

\noindent where $\bar{s}= (\gamma+1)M^{2}
/(\gamma+M_{0}^{2}(\gamma-1))$ 
and $\bar{\Delta}=(4M_{0}^2-\gamma+1)/(\gamma+1)$ 
are are the density and compression ratios of a 
pure hydrodynamic shock. The angular dependences
obey $f(0) = g(0) = 0$ and $f(\pm\pi/2) = 
g(\pm\pi/2) = 1$. As expected, the shock anisotropy is small if 
$R_M/\beta$ is small. 

{\it Effects of} $\vec{B}$ {\it on radiation} - Even when $\vec{B}_{ext}$
does not alter the shock structure, it can have other effects on SBSL.
We now consider the effects on the SBSL radiation. 
Shock simulations suggest maximum temperatures of
$\sim 10^{8}\,^{\,\circ}$K  \cite{WU} while spectral fits from radiation emanating 
from the bubble center suggest a more
modest $\sim 10^{4\,\circ}$K \cite{SUSLICK}.
The average thermal velocity at these temperatures 
is $< 0.1c$, hence, we will only consider the nonrelativistic 
conditions for observing anisotropic cyclotron radiation in competition with 
isotropic Bremsstrahlung or blackbody emission.  
We ignore collective effects and thus require
$\omega_{B} > \omega_{p}$, the plasma frequency. 
For appropriately dense plasmas, 
$\omega_{p} \approx 10^{11}-10^{12}$/s, requiring $B_{0} > 1-10$T.

If shock asymmetry is negligible ({\it e.g.} if $R_{M} \ll 1$), 
then the plasma feels the externally applied magnetic field
$\vec{B}_{0} \simeq B_{ext}\hat{z}$ and charge trajectories obey
$\dot{\bf v}_e =(e/m_e){\bf v}_e\times\vec{B}_{0}$, where ${\bf v}_e$ is the
electron velocity.  The dipole approximation for cyclotron emission power 
per unit solid angle of a collection 
of uncorrelated electrons is given by 
\cite{JACKSON}
  
\begin{equation}
dL_{c}/d\Omega \simeq (n_e e^2/4\pi c^3)(\hat{n}\times(\hat{n}\times{\dot{\bf v}}_e))^2 = 
(n_{e}e^4/4\pi m_e^2 c^5)v_{\perp}^{2}B_{0}^{2}(1-\sin^{2}\theta\sin^{2}\varphi)
\end{equation}

\noindent where $n_{e}$ is the ionized electron 
density, $\hat{n}$ is in the direction of the observer, $\theta$ 
is the angle between the field and the line of sight, 
and $\varphi$ is the azimuthal angle to be averaged over.
Note that for $\omega_{B}^{-1} \ll 50$ps, also requires
$B_{0} > 1-10$T.

To get the integrated power as a function of $\theta$,
we average over the uncorrelated ${\bf v}_{e}$ using 
$f(v_e,x,t) d^{3}{\bf v}_{e} =  \mbox{exp}[-v_T^{-2}(v_{\perp}^{2}+v_z^2)]
v_{\perp}dv_{\perp}dv_{z} d\varphi$, where $v_T\equiv (2 k_B T/m_e)^{1/2}$.
We also impose a cutoff on the velocity,
$v_{z}, v_{\perp} < \ell_{c}\omega_{B}$, where
$\ell_{c}={\rm min}\{\lambda,R^{*}\}$,
$\lambda = 4k_B^2T^2/n_e e^4$, is the mean free path for electron
collisions, and $R^{*}$ is the size of the ionized region. This ensures that the charge can
coherently twist around the magnetic field sufficiently to 
radiate near $\omega_{B}$. The cyclotron power is then

\begin{equation}
L_{c}(\theta) = n_{e} (e^{4}B_{0}^2v_T^2 /8\pi m_e^2 c^5)
\left[1-(\Lambda+1)e^{-\Lambda}
\right]\mbox{Erf}(\sqrt{\Lambda})(1+\cos^{2}\theta),
\label{Lc1}
\end{equation}

\noindent where $\Lambda \equiv 
(\ell_{c}\omega_{B})^{2} m/2k_{B}T$.  When $\ell_c=\lambda < R_c$, 
$\Lambda\propto T^3$, but when $\ell_c= R_c < \lambda$, $\Lambda \propto 1/T$.
Thus the maximum range of $v_\perp,v_z$ contributing to $L_{c}$ would occur
where $\lambda\simeq R_c$.  When $\Lambda \gg 1$, 
$L_{c}(\theta) = n_{e}(e^4v_T^2B_0^2/16\pi m_e^2 c^5)
(1+\cos^2\theta)$, the standard expression \cite{BOYD}.  
When $\Lambda \ll 1$, 

\begin{equation}
L_{c}(\theta) \simeq n_e(v_T^2\pi)^{-3/2}
\left({e^{2}\over 8c}\right)\ell_{c}^{5}\omega_{B}^{7}
(1+\cos^{2}\theta) + O(\Lambda^{7/2}).
\label{Lc3}
\end{equation}

To determine when anisotropic emission may be observable,
we compare the cyclotron intensities with those from 
isotropic Bremsstrahlung and blackbody emissions. 
We consider a singly ionized species in a neutral
plasma, justified because the Debye length $\ll R^{*}$.
Near $\omega_{B}$, the isotropic Bremsstrahlung radiation power  
in the dipole approximation is,

\begin{equation}
L_{brem}(\omega_{B}) \simeq 
\int_{\omega_{B}-\Gamma/2}^{\omega_{B}+\Gamma/2}\! d\omega\,
{dL_{brem}(\omega)\over d\omega} \simeq 
{16n_{e}^{2}e^{6} \over 3 c^{3}m_e^{2}}\!
(v_T^2\pi)^{-1/2}\!
\ell n \left({v_T ^2 \over \omega^{2}b_{min}^{2}}\right)
\Gamma\label{Lbrem}
\end{equation}

\noindent where $b_{min}\sim 4e^2/\pi m_e v_T^2$ is a 
minimum impact parameter, and 
where $\Gamma$ is the cyclotron  line width 
or the narrow bandwidth of a detector.  
Detector bandwidths can be made very narrow ($< 1{\rm cm^{-1}}$), so we use 
$\Gamma \simeq \omega_{col}\sim (2k_bT/m_e\lambda^2)^{1/2}\sim 7\times10^{12} (n_e/10^{22})
(T/10^6{\rm K})^{-3/2}{\rm s^{-1}}$ from collisional 
broadening, which dominates Doppler broadening for $T< 10^8$K.
(We have assumed the neutral plasma to have 
a density  $\sim 10^7$kg/m$^3$ 
consisting primarily of singly ionized nitrogen.)
The power radiated by a blackbody at
low frequencies (for $B = 10$T, $\omega_{B}$ is
in the IR/microwave region $ \simeq 1.8 \times 10^{12}$/s) is 
$L_{bb}(\omega_{B}) \simeq 
\Gamma  \omega_{B}^{2}k_{B}T/( \pi^2 c^{3})\sim
10^{-3}(\omega_B/1.3\times 10^{13}{\rm sec^{-1}})^2
(T/10^6{\rm K})(\Gamma /7\times 10^{12}{\rm sec^{-1}}){\rm erg/sec}$. 
Similarly,  $L_{brem}\sim 6.4\times 10^{14}
(n_e/10^{22}{\rm cm^{-3}})^2(T/10^6{\rm K})^{-1/2}
(\Gamma/7\times10^{12}{\rm sec^{-1}}){\rm erg/sec}$, where
the Gaunt factor logarithm in (\ref{Lbrem}) is $\sim 1$.
Now $\Lambda\sim 10^{-4}$, for these characteristic parameters, and 
$L_{c}\sim 6\times  10^{26}(n_e/10^{22}{\rm cm^{-3}})
(\omega_B/1.3\times 10^{13}{\rm sec^{-1}})^7
(\ell_c/10^{-6}{\rm cm})^5(T/10^6{\rm K})^{-3/2}$erg/sec
so $L_{c}$ will dominate.  In this estimate we have taken
$\ell_c\sim R^{*}$.  Note that the dominance of
$L_{c}$ depends strongly on $\omega_B$.

{\it Discussion} - We have illustrated how a 
magnetic field $\vec{B}_{ext}$ may break the 
symmetry of a collapsing ionized region or 
a propagating shock (Fig. \ref{ASYM}) in two limits
of $R_M$. Using Spitzer's formula to estimate 
$\sigma_{p}$ at $10^{8\,\circ}$K and typical 
velocity and length scales from shock simulations \cite{WU},
we conservatively estimate $R_{M} \approx 10^{-3}$. 
However, in general, consider Guderly's solution \cite{WHITHAM} for 
a spherical hydrodynamic shock:
$r(t) \propto t^{n}$, where $t=0$ is the time the imploding shock 
collapses. Here, $v(t) \propto t^{n-1}$,
and $R_M \simeq \sigma_{p}t^{2n -1}$. For noninteracting
gases, $n>1/2$, implying $R_{M}\rightarrow 0$ as the shock converges 
($t\rightarrow 0$) and rebounds. It is possible that $n<1/2$ for
compressed materials with large $\gamma = C_{p}/C_{v}$. 
Experiments using gases that have $n<1/2$ when 
compressed and heated, if they exist, would be in 
the $R_{M}\rightarrow \infty$ regime, where shocks will be strongly
perturbed by $\vec{B}_{ext}$ (Fig. \ref{ASYM}). 

In either case, large or small $R_{M}$, 
anisotropies develop during the evolution of a shock. 
These anisotropies are greater for 
smaller $\beta$ (larger $B_{0}$), but vanish as 
$R_{M}/\beta \rightarrow 0$.
Spherically converging hydrodynamic shocks are inherently unstable. 
If luminescence requires a spherically converging shock, $\vec{B}_{ext}$ 
may further destabilize the shock leading to reduced heating 
upon implosion and destroy luminescence. 
Comparison of experiments with numerical 
results can then produce an estimate of $R_{M}$.
Furthermore, the $\theta$-dependent  pressure exerted on
the liquid/gas interface by a recolliding shock 
may destabilize the nonlinear bubble oscillations. 
An external magnetic field can thus indirectly alter the 
region of bubble oscillation stability \cite{BRENNER}.
It is thought that bubble stability is crucial for SBSL. 
Experiments with external fields can determine the 
relative importance of symmetric shocks for 
radiation phenomenon and stable bubble oscillations in SBSL. 
Differences between multiple bubble sonoluminescence
(MBSL) and SBSL can also be probed;  if only SBSL depends 
strongly on symmetric bubble oscillations,  
an external field would affect SBSL and not MBSL.

For low $R_{M}$, especially with strong magnetic 
fields, the radiation can be expected to have an
anisotropic cyclotron component given by (\ref{Lc1}). 
Experimental determination of the
luminescing spectrum is interrupted by the
surrounding water which cuts out light with
wavelength below $\sim$ 220 nm.  We therefore suggest that
structure may be observed at {\it longer} IR and microwave 
wavelengths, near $\omega_{B}$ and its higher harmonics. 
An angular dependence of the form $\sim 1+\cos^{2}\theta$ may 
be observed if enough signal can be collected from 
a narrow bandwidth detector set at $n\omega_{B}$. 
Given the stable, oscillatory nature of SBSL, this seems feasible.
One may be able to discern structure at these harmonics 
by comparing the complicated low frequency spectra
with and without $\vec{B}_{0}$. It is straightforward  
to estimate the cyclotron contribution
by comparing $L_{c}/(L_{c}
+L_{brem}+L_{bb})$ at $\theta = 0$ and $\pi/2$ using 
(\ref{Lc1}) and (\ref{Lbrem}). If no effects 
are observed with strong $\vec{B}_{ext}$, 
the interpretation depends on the optical depth:
If the  plasma were optically thick, blackbody 
would be the only emission mode.
If the plasma were optically thin, the absence of cyclotron emission
means that the velocity cutoff determined by $\Lambda$ would be very small
because  $\omega_{col} \gg \omega_B$. 
$L_{brem}$ would therefore dominate. Since
$L_{brem}$ and $L_{bb}$ differ by many orders of magnitude
near $\omega_B$, the two cases should be distinguishable.
Radiation anisotropy may be detected at low $R_{M}$. 

The spectroscopy of a sonoluminescing bubble in a magnetic field
can help put bounds on relevant physical parameters such as
charge density, and temperature.  
In particular,  $L_{c}\propto T^{-3/2}$ when $\ell_c \simeq  R^{*}< \lambda$,
but  $L_{c}\propto T^{17/2}$ when $\ell_c \simeq \lambda< R^{*}$.
The temperature when $R^{*}=\lambda$, above which 
$L_{c}\propto T^{-3/2}$, is given by
$T_{crit}\sim 10^5 (R_c/10^{-6}{\rm cm})^{1/2}(n_e/10^{22}{\rm cm^{-1}})^{1/2}$K.
Since increasing the acoustic driving pressure increases $T$,
the functional change in $L_{c}$ (anisotropy)  
could help constrain $T$. Finally, note that experiments hitherto have 
time-integrated the radiation. A time dependent spectral analysis of 
asymmetric radiation near $\omega_{B}$ can be a means of monitoring the 
temperature and charge density of the heat-up and cool-down
phases of the bubble oscillation, in addition to the 50ps burst.
Experiments of this nature may help clarify
the mechanisms involved in sonoluminesence. 

T.C. thanks M. Brenner and R.V.E. Lovelace for helpful 
discussions and N. Kannan for assistance in plotting Fig. 2(a).

\end{document}